\documentclass[sigconf,natbib=true,anonymous=false]{acmart}

\usepackage{algorithm}
\usepackage[noend]{algpseudocode}
\usepackage{amsmath}
\usepackage{balance}
\usepackage{caption}
\usepackage{color}
\usepackage{enumitem}
\usepackage{graphicx}
\usepackage{multirow}
\usepackage{subcaption}
\usepackage{url}

  {\begin{list}{}{%
    \settowidth{\labelwidth}{\textbf{#1:}}%
    \setlength{\leftmargin}{\labelwidth}\addtolength{\leftmargin}{\labelsep}}}%
  {\end{list}}

\AtBeginDocument{%
  \providecommand\BibTeX{{%
    \normalfont B\kern-0.5em{\scshape i\kern-0.25em b}\kern-0.8em\TeX}}}

\copyrightyear{2022}
\acmYear{2022}
\setcopyright{usgovmixed}\acmConference[SIGIR '22]{Proceedings of the 45th International ACM SIGIR Conference on Research and Development in Information Retrieval}{July 11--15, 2022}{Madrid, Spain}
\acmBooktitle{Proceedings of the 45th International ACM SIGIR Conference on Research and Development in Information Retrieval (SIGIR '22), July 11--15, 2022, Madrid, Spain}
\acmPrice{15.00}
\acmDOI{10.1145/3477495.3531991}
\acmISBN{978-1-4503-8732-3/22/07}

\begin{document}

\title{Human Preferences as Dueling Bandits}

\author{Xinyi Yan}
\affiliation{
  \department{School of Computer Science}
  \institution{University of Waterloo}
  \city{Waterloo}
  \state{Ontario}
  \country{Canada}}
\email{xinyi.yan@uwaterloo.ca}
\author{Chengxi Luo}
\affiliation{
  \department{School of Computer Science}
  \institution{University of Waterloo}
    \city{Waterloo}
  \state{Ontario}
  \country{Canada}}
\email{chengxi.luo@uwaterloo.ca}
\author{Charles L. A. Clarke}
\affiliation{
  \department{School of Computer Science}
  \institution{University of Waterloo}
    \city{Waterloo}
  \state{Ontario}
  \country{Canada}}
\email{claclark@gmail.com}
\author{Nick Craswell}
\affiliation{
  \institution{Microsoft}
  \city{Bellevue}
  \state{Washington}
  \country{USA}}
\email{nickcr@microsoft.com}
\author{Ellen M. Voorhees}
\affiliation{
  \institution{National Institute of Standards and Technology}
  \city{Gaithersburg}
  \state{Maryland}
  \country{USA}}
\email{ellen.voorhees@nist.gov}
\author{Pablo Castells}
\affiliation{
  \institution{Universidad Aut\'onoma de Madrid}
  \city{Madrid}
  \country{Spain}
}
\email{pablo.castells@uam.es}

\begin{abstract}
The dramatic improvements in core information retrieval tasks engendered by neural rankers create a need for novel evaluation methods. If every ranker returns highly relevant items in the top ranks, it becomes difficult to recognize meaningful differences between them and to build reusable test collections. Several recent papers explore pairwise preference judgments as an alternative to traditional graded relevance assessments. Rather than viewing items one at a time, assessors view items side-by-side and indicate the one that provides the better response to a query, allowing fine-grained distinctions. If we employ preference judgments to identify the probably best items for each query, we can measure rankers by their ability to place these items as high as possible. We frame the problem of finding best items as a dueling bandits problem. While many papers explore dueling bandits for online ranker evaluation via interleaving, they have not been considered as a framework for offline evaluation via human preference judgments. We review the literature for possible solutions. For human preference judgments, any usable algorithm must tolerate ties, since two items may appear nearly equal to assessors, and it must minimize the number of judgments required for any specific pair, since each such comparison requires an independent assessor. Since the theoretical guarantees provided by most algorithms depend on assumptions that are not satisfied by human preference judgments, we simulate selected algorithms on representative test cases to provide insight into their practical utility. Based on these simulations, one algorithm stands out for its potential. Our simulations suggest modifications to further improve its performance. Using the modified algorithm, we collect over 10,000 preference judgments for pools derived from submissions to the TREC 2021 Deep Learning Track, confirming its suitability. We test the idea of best-item evaluation and suggest ideas for further theoretical and practical progress.
\end{abstract}




\maketitle

\section{Introduction}

Pairwise preference judgments provide an alternative to pointwise graded relevance judgments for the offline evaluation of information retrieval systems~\cite{cbcd08,cvs21,sz20,xie20,hb17a,rcmd22}. When compared to pointwise relevance judgments~---~which assess individual items one by one~--~human assessors can make pairwise preference judgments faster and more accurately~\cite{cbcd08}. By considering two items side-by-side, assessors can recognize fine distinctions between them that may be missed by pointwise judgments~\cite{cvs21,Bah}. Assessment guidelines for pairwise judgments can also be simpler than assessment guidelines for pointwise judgments because they can be expressed in relative terms~\cite{cvs21}. Preferences also bypass the difficult problem of converting ordinal relevance grades into a ratio scale in order to compute NDCG and other standard evaluation measures~\cite{ka09,mmst17,fff21,reviewer0,reviewer1}. 

Recent papers demonstrate the value of preference judgments for offline evaluation. Both \citet{sz20} and \citet{lcs21} define and validate evaluation measures that work directly with a multiset of preference judgments. \citet{avyc21} employ preference judgments to isolate only the best items, measuring rankers by their ability to place these items as high as possible. Others propose and explore measures based on counts of correctly ordered items~--~according to the preference judgments~---~or measures that convert preferences to gain values for use with NDCG and other standard evaluation measures~\cite{sz20,ymt18,xie20,cb08}.

While these papers give considerable attention to evaluation measures based on preferences, they devote relatively less attention to the preference judging process itself. Many papers assume that all pairs in a judging pool will be judged once~\cite{avyc21, cbcd08, sz20}. \citet{sz20} judge all pairs three times, requiring thousands of judgments for a judging pool of only three or four dozen items. \citet{cbcd08} suggest reducing the number of judgments by exploiting transitivity and eliminating non-relevant items in an initial binary relevance pass, but they do not explore these ideas in detail. ~\citet{cvs21} build on these ideas, describing a tournament-like judging strategy targeted at crowdsourced preference judgments.

In this paper, we focus on the judging process, with the goal of identifying the best responses to a query while minimizing the number of preference judgments. We frame this problem in terms of dueling bandits~\cite{yue11,yue12,rucb,mergerucb,bbmh21,mergedts,ll18}. After defining and justifying requirements for human preferences as dueling bandits, we review candidate algorithms from both the machine learning and information retrieval literature. Since the theoretical guarantees provided by these algorithms typically depend on assumptions that are not satisfied by human preference judgments, we simulate selected algorithms to provide insight into their practical utility. Based on these simulations, we propose a modified version of the algorithm described in ~\citet{cvs21} as the most suitable candidate. Using this algorithm, we collect over 10,000 crowdsourced preference judgments for the TREC 2021 Deep Learning Track~\cite{DL21}, verifying the suitability of the algorithm and suggesting future directions for the track.

\section{Dueling Bandits}
\label{sec:require}

The standard (non-dueling) bandits problem is a sequential decision-making process in which a learner chooses one of $K$ ``arms'' at each round and receives real-valued feedback, with the goal of optimizing its evaluation criterion over time~\cite{bandits19}.
The dueling bandits problem represents an important variant in which the
algorithm attempts to identify the probably best of $K$ arms through noisy pairwise comparisons between them~\cite{yue11,yue12,rucb,mergerucb,bbmh21,mergedts,ll18}. While dueling bandits have not been widely employed as a framework for human preference judgments, they have been extensively studied and explored as a tool for ranker selection in the context of online learning to rank~\cite{bbmh21}. \citet{bandits19} provides a recent survey of bandits for information retrieval.

For use with human preference judgments, the requirements for dueling bandit algorithms differ substantially from those for online ranker selection.
We start with a pool of $K$ items, with each item forming an arm. This pool represents a set of items that potentially satisfy a given query. In an academic research context, this pool of items might be formed from experimental runs submitted to an evaluation forum such as TREC. In a commercial context, the pool might be formed from the results returned by production vs.\ experimental rankers. We assume that each item in this pool represents a plausible response to the query, since a preference judgment between items that are unrelated to the query would be pointless. As a result, we assume that the pool has been filtered with a binary relevance pass, eliminating unrelated items.

A preference judgment compares two items $i$ and $j$ from the pool, where $i$ beats $j$ with unknown probability $q_{i,j}$. We assume that each comparison of $i$ and $j$ is independent, with different people making each judgment. We also assume that human judges may be crowdsourced workers or temporary contract employees, with minimal or no training.  As a result, we recognize two sources of noise: 1) genuine disagreements of opinion, where the balance between competing factors might cause independent judges to reach different conclusions; and 2) erroneous judgments due to misunderstandings, distractions, or lack of training. While the use of trained and dedicated assessors might reduce the second source, nothing can be done about the first source.  As a result, even with trained and dedicated assessors we cannot assume that $q_{i,j}$ will typically be close to $0$ or $1$. As an exception, if an assessor is considered to be ``authoritative''~\cite{ow13} then $q_{i,j}$ will be always $0$ or $1$. In this case,  preference judgments are not dueling bandits and traditional sorting and selection algorithms may be applied.

\citet{bbmh21} review common assumptions on pairwise preferences across the research literature. It is common in this literature to define: $\Delta_{i,j}~=~q_{i,j} - 0.5$, and to express the properties and performance of algorithms in terms of $\Delta_{i,j}$ rather than $q_{i,j}$. Experimental evidence reported in prior work suggests that preference judgments by dedicated assessors are very nearly transitive~\cite{cbcd08,sz20}, so that we assume a total order over arms and \emph{strong stochastic transitivity} (SST)~\cite{bbmh21}:
\vspace{-0.5\baselineskip}
\begin{quote}
if  $\Delta_{i,j} \ge 0$ and $\Delta_{j,k} \ge 0$ then
$\Delta_{i,k} \ge \max\left(\Delta_{i,j}, \Delta_{j,k}\right)$.
\end{quote}
\vspace{-0.5\baselineskip}
SST implies other transitivity assumptions listed in \citet{bbmh21}, but not the stochastic triangle inequality.

``No ties'' is another common assumption, and at a purely theoretical level, it might be reasonable to assume that all $|\Delta_{i,j}| > 0$. As long as there is some basis for comparison, i.e., some difference between $i$ and $j$, however trivial, and we have an endless supply of independent human assessors, we would eventually detect a preference difference that cannot be explained by chance. However, practically speaking, any algorithm we employ needs to gracefully handle cases where some values of $\Delta_{i,j}$ are arbitrarily close to zero. The theoretical performance guarantees of some algorithms are expressed in terms of the inverse $1/\Delta_{i,j}$. Rather than automatically excluding these algorithms, we consider the practical impact when $\Delta_{i,j} = 0$ for some pairs. We avoid asking assessors if two items appear equal, or nearly so, since we have no universal way to define ``near equality''. Instead, we require assessors to seek meaningful differences, however minor.

Ambiguous queries provide an example where $\Delta_{i,j}$ can be close to zero, even when $i$ and $j$ represent two of the best items. The first two passages in Figure~\ref{fig:jaffe} are among the top~5 selected for the ambiguous query {\em ``Who is Jaffe?''}. They both provide clear and concise biographies for two different actors with the last name ``Jaffe''. Based on the relative recency and number of their roles, Taliesin, with 170 credits on IMDb, might be preferred to Nicole, who is best known as the voice of Velma on Scooby-Doo. Nonetheless, either is preferable to passages that provide information on the name itself or less focused biographical information. 

Preference judgments are expensive. For the crowdsourced experiments reported in this paper, we spent nearly 36~cents per judgment. As a result, we focus on minimizing the number of comparisons, rather than on other measures of regret. Practically there is also a limit on the number of independent judgments we can obtain for a given pair, since each requires a different assessor.

To summarize, a dueling bandit algorithm for human preference judgments must satisfy the following requirements:
\begin{enumerate}
\item
{\bf Near linearity:} The number of comparisons should grow linearly with $K$, up to a logarithmic term. Strong stochastic transitivity can be assumed to avoid quadratic growth.
\item
{\bf Equality tolerance:} The algorithm must gracefully handle cases where values of $\Delta_{i,j}$ are close or equal to zero. While we do not automatically exclude algorithms whose theoretical performance guarantees include inverse values of $\Delta_{i,j}$, the impact of $\Delta_{i,j} = 0$ values must be considered.
\item
{\bf Parsimony:}
The algorithm must be practically usable, minimizing both the total number of assessments and the number of independent assessments of a given pair. Even with near linear growth and equality tolerance, constant factors can render algorithms too expensive for human assessment.
\end{enumerate}

Our goal is to find the probably best items from a pool of $K$ items.
Given the possibility of ties, {\em Copeland} winners provide a natural definition of ``best'', i.e., the arm or arms that beat the most other arms.
We can also define best items in terms of a sum of expectations, or {\em Borda ranking}, where our goal is to find the items with the greatest {\em Borda score}:
\begin{equation}
b_i~=~\frac{1}{K - 1}\sum_{j = 1, j \ne i}^K q_{i,j}
\end{equation}
Nonetheless, we do not automatically exclude algorithms targeted at other definitions of ``best'', e.g.,
\emph{Condorcet} winners.

\begin{figure*}[t]
\begin{minipage}[c]{0.9\textwidth}
{\bf msmarco\_passage\_50\_318366271}:
Nicole Jaffe (I) Nicole Jaffe. Jaffe's career was mainly based on doing voice-over work. She is best remembered as the voice of the nerdy-and-intellectual Velma Dinkley in the Scooby-Doo cartoons from 1969-1973; before actress Pat Stevens picked up the role.\\
{\bf msmarco\_passage\_30\_571323592}:
Biography. Taliesin Jaffe born as Taliesin Axelrod Jaffe is an American actor, voice actor, ADR director, and scriptwriter. Jaffe was born on January 19, 1977. He is one of the famous director and actor who was born in Los Angeles, California, United States. His nationality is American.\\
{\bf msmarco\_passage\_08\_678686094}:
The Given Name Jaffe. Jaffe is a great choice for parents looking for a more unique name. A name fit for a child full of greatness and distinction, a little adventurer. Although unique, your little daring Jaffe, is sure to make it a memorable one.\\ 
{\bf msmarco\_passage\_24\_359189946}:
Taliesin Jaffe     The Flash. Taking the helm of the legendary Scarlett Speedster is Taliesin Jaffe, who is a video game and television actor. Jaffe has appeared in numerous titles including Fire Emblem Heroes, Final Fantasy XV, Fallout 4, Pillars of Eternity, and Street Fighter IV.
\end{minipage}
\caption{
  Four of 92 pooled passages for the question {\em Who is Jaffe?} (\#1103547).
  The first two were preferred by assessors.
  The last two were randomly selected from the 87 passages not appearing in the top~5.
  Even though the question is ambiguous, passages that provide clear and complete biographies appear to be preferable.}
\label{fig:jaffe}
\end{figure*}

\section{Candidate algorithms}

We review candidate dueling bandit algorithms from the machine learning and information retrieval literature in light of the requirements listed in the previous section. We consider the suitability of these algorithms for identifying the best item or items in the pool through human preference judgments. In undertaking our review, we were greatly assisted by the comprehensive survey on dueling bandits by \citet{bbmh21}, which served as a foundation for our review. 

It is clear that most algorithms do not provide a close match to our requirements for near linearity, equality tolerance and parsimony.  While we do not want to exclude algorithms on purely theoretical grounds, some algorithms employ basic strategies that are clearly inconsistent with human preferences. For example, strategies that repeatedly compare two arms until a winner is found may require too many independent assessors to be
feasible when arms are nearly equal~\cite{falahatgar2017maximum, falahatgar2017maxing, ren2020sample}.

The remaining algorithms could also be excluded on theoretical grounds. Some require quadratic judgments, violating near linearity. Others have performance guarantees expressed in terms of $1/\Delta_{i,j}$, violating equality tolerance. Others assume the stochastic triangle inequality or require the existence of a Condorcet winner, i.e., a single arm that beats all others with probability greater than $0.5$. Although they could be excluded on theoretical grounds, we selected several algorithms for further consideration based on their potential to work in practice, even if their theoretical assumptions are not satisfied. These algorithms all use basic strategies that seem plausible in the context of human preferences.

Beat the Mean~\cite{yue11} and Interleaved Filtering~\cite{yue12} are two of the early dueling bandit algorithms. They both make restrictive assumptions about the underlying preference relations, which do not match our requirements. Later algorithms such as RUCB~\cite{rucb}, MergeRUCB~\cite{mergerucb} and MergeDTS~\cite{mergedts} make weaker assumptions, but still require the existence of a Condorcet winner.
Of these algorithms, we selected MergeDTS for further consideration as representative of the state-of-the-art. MergeDTS is also specifically intended to support a larger number of arms, consistent with the larger pool size seen in some information retrieval evaluation efforts~\cite{DL21}.

Other algorithms do not require a Condorcet winner. Copeland Confidence Bound (CCB)~\cite{zoghi2015copeland}, Scalable Copeland Bandits (SCB)~\cite{zoghi2015copeland} and Double Thompson Sampling (DTS)~\cite{dts} find Copeland winners while minimizing the cumulative regret. Of these, we selected the DTS algorithm for further consideration because experiments have shown that DTS outperforms the other two~\cite{dts}. The Round-Efficient Dueling Bandits algorithm~\cite{ll18} finds a probably best arm with high probability. We selected this algorithm for further consideration, since it is specifically intended to support human tasks such as identifying the best player in sports tournaments.

We also consider candidate algorithms from the literature related to human preferences for information retrieval evaluation~\cite{cbcd08,cvs21,sz20,xie20,hb17a,rcmd22}. While there exists a substantial body of research on both dueling bandits and human preference judgments in the context of information retrieval, explicit connections between them are rare. \citet{bandits19} (Chapter 5) reviews both the use of dueling bandits for online evaluation and the use of (non-dueling) bandits for traditional pointwise relevance judgments~\cite{lpb16} but does not connect the two concepts.

Proposals to base information retrieval evaluations on preference judgments date back to the early 1990s~\cite{fs91,rorvig90,yao95}. Many proposals implicitly assume that all pairs will be judged one or more times~\cite{avyc21, cbcd08, sz20}, with some proposals attempting to reduce the total effort by exploiting transitivity, recognizing ties, and filtering obviously off-topic items~\cite{hb17a,cbcd08}.  \citet{cvs21} and \citet{avyc21} focus on finding the top items from a pool. A recent paper by \citet{rcmd22} considers methods for selecting pairs when the starting point is a set of ranked lists rather than a pool of items. We selected the unnamed algorithm in \citet{cvs21} for further consideration, since it shares our goal of finding top items and accommodates pools with hundreds of items. 

\vspace*{-1\baselineskip}
\subsection{Double Thompson Sampling}

Copeland winners generalize the concept of a Condorcet winner and are guaranteed to exist. \citet{dts} propose Double Thompson Sampling (DTS) to find Copeland winners. The algorithm uses preference probability samples drawn from beta posterior distributions to select candidates for pairwise comparison. When selecting the first arm, DTS excludes the arms that are unlikely to be potential champions based on confidence bounds, and chooses the arm beating most others according to the sampled probabilities. For the second candidate, the posterior distributions are used again to select the arm from uncertain pairs that performs the best compared to the first candidate. The distributions are then updated accordingly based on the comparison results and the procedure is repeated for a time horizon $T$. Under the assumption that there are no ties, DTS achieves a theoretical regret bound of the form $O(K^2 \log T)$, which grows quadratically in the number of arms. While quadratic growth violates our requirements, the algorithm may still be practically suitable for small $K$.

\subsection{Merge Double Thompson Sampling}

Merge Double Thompson Sampling (MergeDTS)~\cite{mergedts} uses a divide and conquer strategy along with Thompson sampling strategies to identify the Condorcet winner. It proceeds as follows: 1) arms are randomly partitioned into disjoint batches of a predefined size. At each time step, MergeDTS selects and compares two arms within one batch to avoid global pairwise comparisons. Similar to DTS, the first candidate is the one that beats most of the others. However in MergeDTS, the second candidate is the first candidate's worst competitor. The rationale behind this choice is to eliminate arms as quickly as possible. An arm is eliminated if it loses to another arm by a wide margin. Once the current batch size becomes one, it is merged with another batch. The entire process iterates until there is only a single-element batch left, which contains the Condorcet winner with high probability.

To provide its theoretical guarantees, MergeDTS makes the following assumptions: 1) a single Condorcet winner exists; 2) there are no ties, unless they are between ``uninformative'' arms that cannot beat any other arms; 3) uninformative arms represent at most one third of the full set of arms. Under these assumptions, MergeDTS provides a high probability bound on the cumulative regret, which is of the order
\[
O\left(\frac{K\log T}{\Delta_{min}^2} \right)
\mbox{,\ where\ } \Delta_{min} = \min\limits_{\Delta_{i,j} > 0} \Delta_{ij}.
\]
\subsection{Round-Efficient Dueling Bandits}

\citet{ll18} study the problem of winner selection in sports tournaments and present an algorithm to find the optimal arm or a Borda winner while minimizing the total number of comparisons. They assume an unknown strength parameter for each arm and a linear relationship between the underlying strength and preference probabilities, so that an arm with higher strength is more likely to win. Given an allowed error probability $\delta$, the algorithm randomly partitions arms into disjoint pairs for comparison, and successively eliminates candidates if they lose to the empirically best arm based on the confidence intervals of the estimated latent strength. The process stops when only a single arm remains or enough iterations have been made. Assuming all arms can be arranged in total order, the algorithm provides a theoretical guarantee of identifying the best arm $i^*$ with probability at least $1 - \delta$ via comparisons at most:
\[
O\left(\Sigma_{j \ne i^*} \frac{1}{\Delta_{i^*, j}^2} \log \frac{K}{\delta \Delta_{i^*, j}}\right).
\]

\begin{figure}[p]
\caption{Function {\bf prefBest}(Pool, $K$, $n$, $m$) returns the probably best item from a pool of $K$ items based on~\citet{cvs21}. The function $\mbox{\bf preference}(u, v)$ returns the item preferred by a human assessor or any other source of preferences.}
\begin{tabbing}
\ \ \ \ \ \ \ \= \ \ \ \ \ \ \ \= \ \ \ \ \ \ \ \= \ \ \ \ \ \ \ \= \ \ \ \ \ \ \ \=\kill
\rule{\columnwidth}{0.4pt}\\
\>{\bf Input}\\
\>\>Pool\>\>Set of items to judge\\
\>\>$K$\>\>Pool size\\
\>\>$n$\>\>Pairings during pruning phase\\
\>\>$m$\>\>Threshold for completion phase\\
\\
\>{\bf Output}\\
\>\>Probably best item in Pool\\
\rule{\columnwidth}{0.4pt}\\
\\
\>{\bf def randomPairings}(Pool, $K$, $n$):\\
\>\>\begin{minipage}[t]{0.80\columnwidth}
Generate and return the edge set of a random graph, where the $K$ items in the pool form the vertices of the graph and each vertex has outdegree $n$ (or $n + 1$).
\end{minipage}\\
\\
\>{\bf def completePairings}(Pool, $K$):\\
\>\>\begin{minipage}[t]{0.80\columnwidth}
Generate and return the edge set of a complete graph, where the $K$ items in the pool form the vertices of the graph (each vertex has outdegree $K -1$).
\end{minipage}\\
\\
\>{\bf def estimateBorda}($E$):\\
\>\>${\bf W} \leftarrow {\bf 0}$\>\>\#{\em count of wins for each item}\\
\>\>${\bf C} \leftarrow {\bf 0}$\>\>\#{\em count of pairings for each item}\\
\>\>{\bf for} ($u$,$v$) {\bf in} $E$:\\
\>\>\>$w \leftarrow \mbox{\bf preference}(u, v)$\\
\>\>\>$C_u \leftarrow C_u + 1$\\
\>\>\>$C_v \leftarrow C_v + 1$\\
\>\>\>$W_w \leftarrow W_w + 1$\\
\>\>{\bf return} ${\bf W}/{\bf C}$\\
\\
\>{\bf def prune}(Pool, $K$, $n$):\\
\>\>$E \leftarrow \mbox{{\bf randomPairings}(Pool, $K$, $n$)}$\\
\>\>$B \leftarrow \mbox{{\bf estimateBorda}($E$)}$\\
\>\>$\mbox{Pruned} \leftarrow \emptyset$\\
\>\>{\bf for} $b_i$ {\bf in} $B$:\\
\>\>\>{\bf if} $b_i \ge 0.5$:\\
\>\>\>\>$\mbox{Pruned} \leftarrow \mbox{Pruned} \cup \{i\}$\\
\>\>{\bf return} Pruned\\
\\
\>{\bf def finalize}(Pool, $K$):\\
\>\>$E \leftarrow \mbox{{\bf completePairings}(Pool, $K$)}$\\
\>\>$B \leftarrow \mbox{{\bf estimateBorda}($E$)}$\\
\>\>{\bf return} $\left\{j~|~b_j = \max_{i \in B} b_i\right\}$\\
\\
\>{\bf def prefBest}(Pool, $K$, $n$, $m$):\\
\>\>$\mbox{\bf while\ } K > m$:\\
\>\>\>$\mbox{Pool} \leftarrow \mbox{{\bf prune}(Pool, K, $n$)}$\\
\>\>\>$K \leftarrow |\mbox{Pool}|$\\
\>\>$\mbox{{\bf return finalize}(Pool, $K$)}$\\
\rule{\columnwidth}{0.4pt}\\
\\
\end{tabbing}
\label{fig:alg}
\end{figure}

\subsection{\citet{cvs21}}
\label{sec:alg:clarke}

While they do not frame their work in terms of dueling bandits, \citet{cvs21} present an algorithm that uses crowdsourced preference judgments to find the top answers to a question from a pool of possible answers.  In that paper, they focused on the potential benefits of preference judgments for information retrieval evaluation, including the impact of preference judgments on the outcomes of TREC 2019 CAsT retrieval experiment. However, they did not explore the algorithm itself in any detail. Here, we consider the algorithm as a solution to the dueling bandits problem.

Figure~\ref{fig:alg} presents a version of the algorithm modified to return a set of best responses. We base this presentation on the paper itself, but confirm it against their implementation. The algorithm works in a series of pruning phases. During each phase, the algorithm estimates Borda scores for the remaining items in the pool and prunes those items with estimated Borda scores less than or equal to $0.5$. Pruning continues until the pool size drops to a threshold $m$, at which point the algorithm makes a final estimate of Borda scores and returns the set of items with the greatest scores.

When the size of the pool is greater than $m$, the algorithm estimates Borda scores by first generating a random graph in which each item is randomly paired with at least $n$ other items, with no pairing repeated. If the size of the pool is odd, one item will be paired with $n + 1$ others. In their implementation, ~\citet{cvs21} use a brute force approach to generate this graph. Assessors judge the pairs and the algorithm estimates Borda scores from the counts of pairs won and lost. While \citet{cvs21} provide no theoretical guarantees, their algorithm produces crowdsourcing results they show to be consistent with the judgments of dedicated assessors. Their overall approach is similar to the work of \citet{karnin13}, which solves the equivalent problem for (non-dueling) bandits in both fixed confidence and fixed budget settings, and that paper might provide a starting point for further theoretical investigation.

\section{Simulations}
\label{sec:sim}

To better understand the ability of these algorithms to satisfy the requirements of Section~\ref{sec:require}, we simulate them on two artificial test cases. One test case (Case A) assumes a total order of items with no ties. We would hope that this case could be handled easily by most dueling bandit algorithms since it satisfies the most common requirements of these algorithms. The other test case (Case B) represents something close to the opposite extreme, with many ties and no single winner. Both cases assume $K = 100$, with items $0, 1, 2, ... 99$:
\begin{tabbing}
\ \ \ \ \ \ \ \= \ \ \ \ \ \ \ \= \ \ \ \ \ \ \ \= \ \ \ \ \ \ \ \= \ \ \ \ \ \ \ \=\kill
\>{\bf Case A} (total order, no ties):\\
\>\>$i < j \implies q_{i,j} = 0.75 \mbox{\ and\ } q_{j,i} = 0.25$\\
\\
\>{\bf Case B} (two winners, many ties):\\
\>\>$q_{0,1} = q_{1,0} = 0.5$\\
\>\>$i > 1 \implies q_{0,i} = 0.75 \mbox{\ and\ } q_{i,0} = 0.25$\\
\>\>$i > 1 \implies q_{1,i} = 0.75 \mbox{\ and\ } q_{i,1} = 0.25$\\
\>\>$i > 1 \mbox{\ and\ } j > 1 \implies q_{i,j} = 0.5$
\end{tabbing}
We picked these test cases for their simplicity and as a contrast to one another. The value of $K = 100$ is close to the limit for the number of arms in the experiments of  Section~\ref{sec:human}. We pick $0.75$ as a ``winning'' probability, since it corresponds to the level at which we tolerate test case errors in our experiments with crowdsourced assessors (Section~\ref{sec:human}). For algorithms that require a time horizon as a parameter, corresponding to a limit on the number of preference judgments, we choose $T = 1000$ as a practical limit on the number of judgments for $K = 100$ items. In any case, we treat $T = 1000$ as a fixed budget and we halt simulations after $1,000$ comparisons. Our simulations use the original code when authors have made it available, maintaining the default values of parameters in that code. When code is not available, we base simulations on our own implementations from pseudo-code and descriptions in the papers. 

We executed each of the selected algorithms $1,000$ times on each of the two cases. We summarize the results in Figure~\ref{fig:sim}. We provide simulation details and a discussion for the individual algorithms in the subsections that follow. Except for ~\citet{cvs21}, algorithms ran until their budget for comparisons was exhausted. The number of comparisons required for~\citet{cvs21} varied depending on the simulation, and we report a range. We determined the maximum number of independent assessors according to the number of times the simulation requested a judgment for each specific pair. The maximum number of assessors also varied from simulation to simulation, and we report a range.

All algorithms performed fairly reasonably on Case~B, with between 63.9\% and 81.4\% of simulations finding at least one or two of the ``winners''.  All algorithms performed less reasonably on Case~A, which we had assumed would be the easier case, with between 10.8\% and 51.0\% finding the single ``winner''. Based on our consideration of the theoretical aspects and these simulations, we can find no dueling bandit algorithm suitable for human preference judgments, beyond the heuristic approach of ~\citet{cvs21}, which lacks an explicit theoretical foundation. Dueling bandit algorithms are sensitive to parameterization. We made efforts to adapt the other algorithms by adjusting parameters, but without success.

\begin{figure*}[t]
\begin{tabular}{r|ccc|cccc}
& \multicolumn{3}{c|}{{\bf Case A}} & \multicolumn{4}{c}{{\bf Case B}}\\
& \multicolumn{3}{c|}{total order, no ties} & \multicolumn{4}{c}{two winners, many ties}\\
{\bf Algorithm}  & best found & comparisons & assessors & one found & both found & comparisons & assessors\\
\hline
Double Thompson Sampling~\cite{dts} & 124  & 1000 & 4-24 & 710 & 7 & 1000 & 4-20 \\
MergeDTS~\cite{mergedts} & 108 & 1000 & 31-93 & 581 & 58 & 1000 & 28-81\\
Round-Efficient Dueling Bandits~\cite{ll18} & 141 & 1000 & 2-5 & 680 & 89 & 1000 & 2-5\\
\citet{cvs21}
& 502 & 599-759 & 2-5 & 666 & 94 & 592-764 & 2-5\\
\citet{cvs21} (extra final phase)
& 510 & 624-781 & 3-6 & 733 & 81 & 616-795 & 3-6\\
\end{tabular}
\caption{Summary of $1,000$ simulations of selected algorithms on two artificial test cases intended to be representative of two possible extremes. Section~\ref{sec:sim} provides details of the simulations and an interpretation of the outcomes.}
\label{fig:sim}
\end{figure*}

\subsection{Double Thompson Sampling}
\citet{mergedts}\footnote{\url{github.com/chang-li/MergeDTS}} provide implementations and optimal parameters for both DTS and MergeDTS. We modified the return statement in their code to return the empirically best arm when the number of comparisons reached 1000. In our simulations, we set $\alpha = 0.8^7$, which they report as the empirically optimal parameter for DTS \cite{mergedts}.
For Case A, 959 simulations returned exactly one item, out of which 124 simulations correctly identified item~0. In addition, 922 items that were not item 0 were returned. For Case B, 710 simulations returned one of the correct answers, and 7 simulations found both winners. Across all simulations, 357 items that were neither item~0 nor item~1 were returned.

\subsection{MergeDTS}
\citet{mergedts} suggest the optimal parameters for MergeDTS are $\alpha = 0.8^6$, $M = 16$ and $C=4,000,000$, and we follow this setting in our simulations. Among~1000 simulations for Case A, MergeDTS found the correct target 108 times and returned 1503 items that were not item 0. For Case B, 581 simulations returned one of the two winners while 58 simulations found both. Across all simulations, 914 items that were neither item 0 nor item 1 were returned.

\subsection{Round-Efficient Dueling Bandits}
We implement the Round-Efficient Dueling Bandits of finding the optimal arm based on the pseudo-code in \citet{ll18}, which we will provide after acceptance\footnote{\url{github.com/XinyiYan/duelingBandits}}. In the simulations, we set the error probability $\delta=0.2$. Of 1000 simulations for Case A, the algorithm returned one item 559 times, where 141 times it found the optimal arm; 1632 items that were not item 0 were returned. For Case B, 680 simulations returned one of the two optimal arms, and 89 simulations found both arms; 606 items that were neither item 0 nor item 1 were returned during the simulations for Case B.

\subsection{\citet{cvs21}}

We use the implementation of \citet{cvs21}\footnote{\url{github.com/claclark/preferences}} for these simulations. We run the code ``as is'', replacing their example judging script with scripts that implement Case~A and Case~B. Since the algorithm is intended for precisely our scenario, we made no changes to the implementation.

For both cases, nearly half of the simulations returned two or more items tied for best~---~497~for Case A and 489~for Case~B. For Case~A, 502~simulations of the 1000~simulations correctly returned item~0, while 323~simulations returned item~1 and 213 simulations~returned item~2. In total, the simulations for Case~A returned 995~items that were not item~0. For Case~B, 94~simulations returned both item~0 and item~1, while 666~returned one or the other, but not both. In addition, the simulations for Case~B returned 729~other items, which were neither item~0 nor item~1.

Given the large number of ties produced by the algorithm ``as is'', we attempted to address the problem by adding an extra final phase. This extra final phase judges all pairs in the final pool a second time so that the algorithm bases its final estimates on two comparisons of each pair. From a practical standpoint, an extra final phase requires only an extra independent assessment for each pair.

With an extra final phase, there are fewer ties and incorrect results. For Case~A, where there is a single best item, 290~simulations return one or more items tied for best. Of the 1000~simulations, 510~correctly returned item~0, while 302~simulations returned item~1 and 166 simulations~returned item~2. In total, the simulations for Case~A returned 780~items that were not item~0. For Case~B, 81~simulations returned both item~0 and item~1, while 733~returned one or the other, but not both. In addition, the simulations for Case~B returned 430~other items, which were neither item~0 nor item~1. 

\section{Human preference judgments}
\label{sec:human}

In our original plan for this research, we expected to experimentally compare several methods from the literature. However, based on our review and simulations, only the algorithm of ~\citet{cvs21} provides a feasible solution to the requirements of Section~\ref{sec:require}. With only one candidate remaining, we focus our experiments on further validation of the algorithm, on opportunities for improvement, and on exploration of sparse labels for information retrieval evaluation, providing support for ideas proposed by~\citet{avyc21}.

\subsection{Method}

We applied the algorithm of ~\citet{cvs21} to crowdsource preference judgments for pools generated from runs submitted to the passage retrieval task of the TREC 2021 Deep Learning Track~\cite{DL21}.
For these experiments we worked with an initial pool of 50 queries and provided by the track organizers. During their own judging process they added seven queries and dropped four, so that the overlap is 46~queries~\cite{TREC21}.  Participants in the track submitted 63~runs to the passage retrieval task. Each run comprised a ranked list of up to 100~passages intended to answer each of 477~questions. TREC assessors judged  these queries on a four-point relevance scale (``perfect'', ``highly relevant'', ``related'', ``irrelevant'')~\cite{TREC21}. The query in Figure~\ref{fig:jaffe} (\#1103547) was one of the dropped query because it has no highly relevant or perfect passages.

We based our experiments on the implementation provided by ~\citet{cvs21} in the associated repository\footnote{\url{github.com/claclark/preferences}}. Apart from the addition of a second finalization phase, we followed the crowdsourcing procedure described in that paper as closely as possible, using their code without change, including default parameters. We implemented the second finalize phase by requesting two judgments for each pair in the final pool through the crowdsource platform, computing scores in a separate script from both the judgments in the final phase of the original implementation and our extra finalization phase. Apart from this additional finalization phase, our procedure is exactly \mbox{{\bf prefBest}(Pool, $K$, 7, 9)} as listed in Figure~\ref{fig:alg}, where $n = 7$ and $m = 9$ are the default values in the original implementation.

\begin{figure}[tp]
\includegraphics[width=3.10in, keepaspectratio]{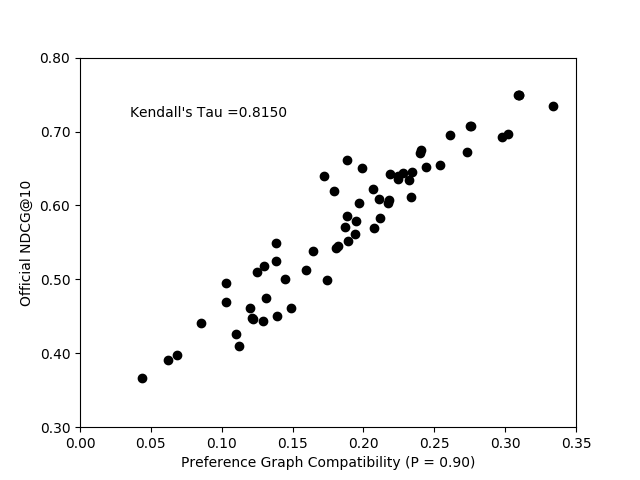}\\
\caption{
Scatter plot comparing experimental runs submitted to the passage retrieval task of the TREC 2021 Deep Learning track on pointwise graded judgments vs.\ pairwise preference judgments. Each point represents a single run.
Official TREC runs use slightly different questions and relevance judgments than our initial pools~\cite{TREC21}.
}
\label{fig:trec}
\end{figure}

We started with depth-10 pools generated from the runs submitted for the passage retrieval task, along with their pointwise graded judgments. The implementation builds its own internal pools from the three grades of relevant passages, omitting irrelevant passages, ``top down'' until a minimum threshold is reached or until all relevant passages are included. For our experiments, we maintained the default value of~5 for this threshold. The implementation first adds all perfectly relevant passages to its pool, if the threshold is not reached, it then adds all highly relevant and then all relevant passages until the threshold is reached or there are no relevant passages remaining. For the 50~questions, preference judging pools ranged in size from~5 to~130, with a median size of~15 and a mean size of 31.4.

We used Amazon Mechanical Turk (MT) to crowdsource our judgments. We used the judging interface described in ~\citet{cvs21} as our starting point, modifying it to simplify data management and improve performance. Each judging task comprises ten target pairs~---~for which we need judgments~---~and three test pairs. These test pairs were intended to reduce judging errors due to misunderstandings, distractions, or the absence of assessor training, as discussed in Section~\ref{sec:require}. Each test pair consisted of a question, a ``best-known answer'' passage, and an off-topic passage. These questions and best-known answers were taken from the collection developed by ~\citet{cvs21}, as provided in their github repository. Target pairs and test pairs were randomly assigned to tasks. Within each pair, the left and right placement of passages was determined randomly. For judging purposes, we edited some of the 50~questions to correct obvious grammatical and other errors. We included these edited questions in the data release for this paper. 

We required crowdsourced workers to have a task approval rate greater than 95\%, and to have more than 1,000 approved tasks. We required workers to be located in the United States, consistent with the context of these TREC experiments. We paid workers \$2.00 for completing each task, consistent with the expected time to complete a single task and our local minimum wage. In addition, we paid a platform fee of \$0.40 to Amazon MT for each task. Under our research protocol, workers could stop after partially completing a task and receive a prorated amount, but none did. While we were debugging the performance of our system a small number of workers experienced unacceptable delays and were paid compensation. Our research protocol received ethics approval from the institution of the lead author. Under this protocol, judgments collected can be released without personal identifiable information, which we commit to do after acceptance of this paper.

Data from workers who did not correctly answer at least 75\% of the test pairs was excluded. This requirement means that workers completing a single task were required to correctly answer all three test pairs; workers completing two tasks were required to correctly answer at least five of the six test questions. We paid these workers for their work, but they were excluded from participating in later phases. 

The total cost for all judgments was \$3,830.41, including interface debugging tasks, excluded data, and platform fees. This produced a total of 10,697 usable judgments at a cost of just under 36~cents each. Of these, 9,713~judgments were required by the algorithm in Section~\ref{sec:alg:clarke} and an additional 984~judgments were required for our extra finalization phase.

\begin{figure*}[tp]
\includegraphics[width=3.10in, keepaspectratio]{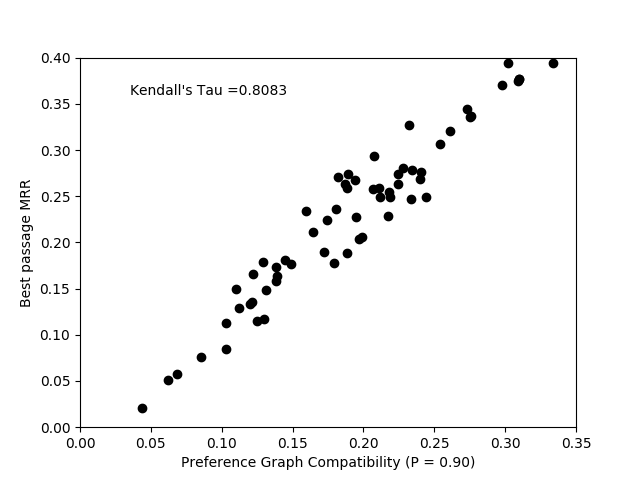}
\includegraphics[width=3.10in, keepaspectratio]{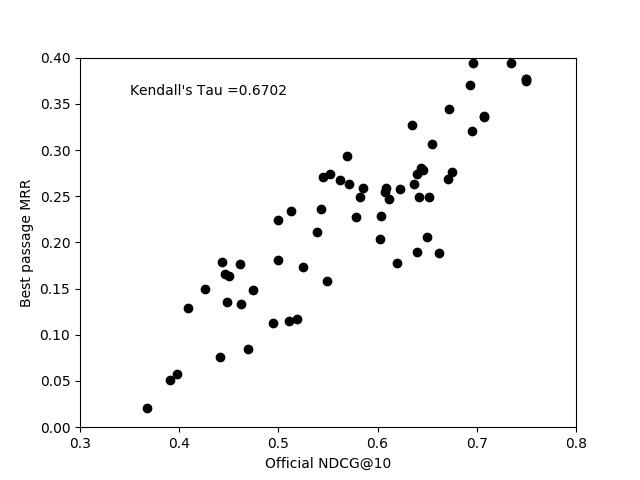}
\includegraphics[width=3.10in, keepaspectratio]{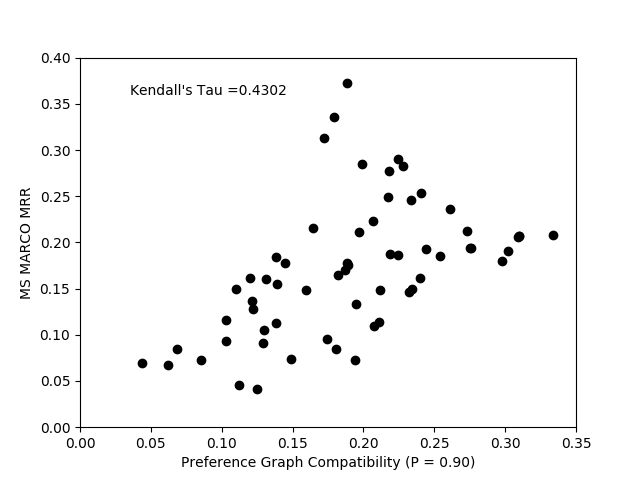}
\includegraphics[width=3.10in, keepaspectratio]{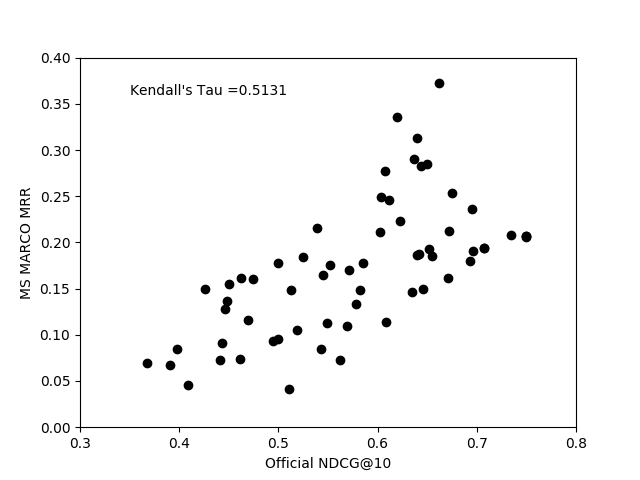}
\caption{
Scatter plots comparing sparse binary judgments against graded judgments and preference judgments for experiments runs submitted to the passage retrieval task of the TREC 2021 Deep Learning Track. Each point represents a single run. Official TREC runs use slightly different questions and relevance judgments than our initial pools~\cite{TREC21}.
}
\label{fig:best}
\end{figure*}

\vspace*{-\baselineskip}
\subsection{Exact duplicates}

After running the entire experiment, including the extra finalization phase, we discovered that the test collection used for the TREC Deep Learning passage retrieval task (the MS MARCO v2 dataset) contained exact duplicates~--~passages that were character-for-character identical. Of the 1570~passages in our combined judging pools, 460~passages contained at least one duplicate in the same pool. If we keep just one of these duplicates, it would leave 1308~items in the pool.

If we had encountered these duplicates earlier in the experiment, it is likely that we would have merged them into equivalence classes, so that assessors would not have to choose between absolutely identical passages. However, we did not notice them until we were looking at top results to select examples for this paper and realized that the top~5 of question \#364210 included an identical copy of the top answer, which was tied for second. We did not see any identical pairs during our interface tests, perhaps because comparisons between exact duplicates are relatively rare, despite the relatively large number of them. Excluding the extra finalization phase, only 169 (1.7\%) of the 9,713~judgments required for our experiments are between identical elements. 

As a result, we have an unplanned test of our equality tolerance requirement. If the top result for a question has an exact duplicate, it should also appear in the top results. Comparing the top results from the main experiment (excluding the extra finalization phase) the top document for 15~questions had a duplicate in the pool. If we exclude five questions where the pool size is less or equal to $m = 9$, where there were no pruning phases, the top document for ten questions had at least one exact duplicate in the pool. The pool sizes for these questions range from~11 to~130, with an average size of~50.5.

For eight of these ten questions, the top passage had a single exact duplicate in the pool. For six of these eight questions, the duplicate also appeared in the top-5, usually ranked second or third. For one of the eight questions, the duplicate reached the upper half but did not reach the finalization phase. For the remaining question, the duplicate was eliminated during the initial pruning phase.

For one question (\#629937) the top passage had two duplicates in the pool, which were ranked third and fourth. One question (\#508292) had two duplicate passages tied for first. These passages had two other duplicates in their pool of 103~passages. One of these duplicates was ranked second. The other made it into the top half but was eliminated during the second pruning phase.

Overall, 9~of~12 duplicates for these ten questions (75\%) reached the top-5, a result consistent with our expectations from the analysis and simulations in previous sections. This outcome might be improved by increasing the value of $n$ to decrease the number of top results lost during the pruning phases. As future work, we hope to extend our theoretical understanding of the approach to choose the number of comparisons during the pruning phase more dynamically, based on the data itself~\cite{karnin13}.

\subsection{Extra finalization phase}
\label{sec:final}

Our simulations suggested that an extra finalization phase would reduce the number of ties and incorrect results. To test this observation, for each question we requested a second round of assessments for the pairs in the final phase. This extra finalization phase gives us three sets of best items. Set~I from the first round, Set~II from the second round, and a Combined~Set constructed by merging the judgments from both finalization phases and computing scores from the combination.

For the 50~questions, both Set~I and Set~II had 75~best passages~---~an average of 1.5~passages per question. The Combined~Set had fewer ties, with 66~best passages~--~an average of 1.32~passages per question. While the exact distributions of ties differ between Set~I and Set~II, both have (different) questions with 4 passages tied for best, while the Combined~Set has none.

Comparing the top passages in Set~I and Set~II, only 22 of the topics share at least one best answer, and the intersection of Set~I and Set~II contains only 25 passages. Even though the Combined~Set is smaller, 40~of the topics in the Combined~Set share a top answer with Set~I; 39~of the topic in the Combined~Set share a top answer with Set~II. The Combined~Set shares 44 passages with both Set~I and Set~II.

Although it is perhaps not surprising, the Combined~Set has fewer ties and is more consistent with both Set~I and Set~II than they are with each other. We report the experiments in the next section on the Combined~Set. As future work, we hope to discover and validate more a principled method for the finalization phase.

\subsection{Best item evaluation}

Traditional approaches to information retrieval evaluation aim for judgments to be as complete as possible. Experimental runs are pooled as deeply as possible; efforts are made to judge as many items as possible. The goal is to make the collection re-usable, so that future experiments conducted on the collection provide meaningful results, even if these new experimental runs surface unjudged documents in their top ranks\cite{bpref, incomplete}. TREC has struggled with this problem of completeness for most of its history, with the overview paper for TREC 2021 providing an excellent summary~\cite{TREC21}. Some TREC tracks, including the Deep Learning Track, use active learning methods~\cite{hical} to expand the set of known relevant documents. Unfortunately, despite these efforts, ``Most topics have very many relevant...and as a result the collection is likely not a reusable collection.''~\cite{TREC21}.

The MS MARCO leaderboards\footnote{\url{microsoft.github.io/msmarco/}} represent a radical departure from traditional information retrieval evaluation methodology. Over the past three years, the MS MARCO leaderboards have tracked dramatic improvements in core information retrieval tasks, including passage retrieval~\cite{marco21full}. MS MARCO bases evaluation on sparse labels, with 94\% of the questions in the development set for passage ranking having only a single known relevant passage~\cite{avyc21}. Given this sparsity of relevance labels, mean reciprocal rank (MRR) forms the primary evaluation measure. While the sparse labels are not intended to represent the ``best'' passages~\cite{DL21}, better systems will tend to place these passages higher in their rankings. This overall approach has been validated through comparisons with Deep Learning Track results in 2019 and 2020~\cite{marco21full}.

\citet{avyc21} suggest that if the sparse labels represented the best known items, the long term validity of the approach might be further improved.
We further suggest that there may be fewer concerns about the re-usability of a collection based on best known items. If a new experiment ranks an unjudged item above the best known item, the unjudged item only invalidates measures if it is better than all previously judged items, a much higher bar than mere relevance.

To further explore the validity of best answer evaluation, we compare best item evaluation with measures computed over all available judgments. For our comparison with graded judgments, we use the official TREC NDCG@10 score. We also compare with Preference Graph Compatibility (PGC)~\cite{lcs21}, which is computed directly from a multi-graph of preferences. Figure~\ref{fig:trec} plots NDCG against PGC, computed over the 63~runs submitted to the passage retrieval task of the TREC 2021 Deep Learning Track~\cite{DL21}.  We compute PGC using the entirety of the preference judgments collected by our experiments, including the extra finalization phase.

The top plots in Figure~\ref{fig:best} are based on the Combined~Set of preference labels from Section~\ref{sec:final}. Despite the sparsity of these preference labels, the ordering of runs is correlated with PGC, with a Kendall's $\tau$ over $0.8$. Both of the top plots show a generally consistent ordering of runs.

Questions for the TREC 2021 Deep Learning Track were taken from questions held out from MS MARCO data releases~\cite{DL21}, so that MS MARCO sparse labels are available for these questions. ~\citet{DL21} report MRR values using these labels, which we use to create the bottom plots in Figure~\ref{fig:best}. With a Kendall's $\tau$ under $0.5$, MRR on these labels is much less correlated with our preference judgments. The top run on the MS MARCO labels (\verb|p_f10_mdt53b|) places just below the median on PGC. The group submitting this run is known for it experience and expertise on the MS MARCO collection~\cite{h2oloo}, expertise which may have inadvertently engendered poorer performance on the preference-based labels.

As a final experiment, we created random sparse labels based on the official graded relevance judgments for the Deep Learning Track. For each of the 50~questions, we pooled the passages with the highest relevance grade for that question. From each pool, we picked a single random relevance passage as the sparse label for that question, and computed MRR for each run using these sparse labels. We repeated this process 1,000 times. For 95\% of these trials, the Kendall's $\tau$ correlation between MRR and NDCG@10 fell between $0.485$ and $0.735$. Kendall's $\tau$ with the MS~MARCO labels ($0.5131$) is closer to the lower end of this range, while Kendall's $\tau$ with the preference labels is closer to the upper end. On the 1,000 trials, the best run achieved an MRR value between $0.120$ and $0.219$. On the MS MARCO labels \verb|p_f10_mdt53b| achieves an MRR of $0.3728$, while on the preference labels \verb|NLE_P_v1| achieves $0.3946$, both well above the best MRR values on random labels. This outcome suggests that neither the MS MARCO labels nor the preference labels are arbitrary relevant passages.
In general, these passages are preferred by the rankers themselves, where rankers heavily tuned on MS MARCO appear to prefer MS MARCO passages. 

\section{Conclusion}

Neural rankers push the limits of traditional information retrieval evaluation methodologies. If most items returned by a ranker are relevant in the traditional sense, how do we measure further improvements? A possible answer is to focus on the top items, the best of the best~\cite{avyc21}. A better ranker would place these top items higher in its ranking. We explore dueling bandits as a framework for finding these items.

After reviewing dueling bandit algorithms in the research literature, we planned to experimentally compare selected algorithms with crowdsourced judgments. After defining requirements in Section~\ref{sec:require}, we determined that only ~\citet{cvs21} fully satisfied them. On the simple test cases of Section~\ref{sec:sim} the practical performance of other algorithms does not warrant further consideration. While many dueling bandit algorithms have proven their practical value in the context of online ranker selection, they have not been applied in the context of human preference judgments, and do not provide a good match for the requirements. As an outcome of our review and simulations, we outline a number of directions for theoretical and practical progress.

Using the algorithm of \citet{cvs21}, we crowdsourced preference judgments for the passage retrieval task of the TREC 2021 Deep Learning Track, identifying the probably best answer for each of the 50~question in the evaluation set. These experiments suggest that the algorithm meets our requirements, particularly the requirement that ties be tolerated. We demonstrate the practical value of an extra finalization phase. These experiments also suggest that evaluation can be based on the best known items alone, supporting the views of \citet{avyc21}.

Code, crowdsourced preference judgments and best answers are available in an associated repository:
\begin{quote}
    \url{https::/github.com/XinyiYan/duelingBandits}
\end{quote}

\section*{Acknowledgements}

We thank Mark Smucker, Gautam Kamath, and Ben Carterette for their feedback.
This research was supported by the Natural Science and Engineering Research Council of Canada through its Discovery Grants program.

\bibliographystyle{ACM-Reference-Format}
\balance
\bibliography{dueling}

\end{document}